\documentclass[journal]{IEEEtran}

\usepackage{amsmath,amssymb,bm,amsthm}
\usepackage{booktabs}
\usepackage{cite}
\usepackage{graphicx}
\usepackage[caption=false,font=footnotesize]{subfig}
\usepackage{multirow}
\usepackage{url}
\usepackage{comment}
\usepackage{balance}
\usepackage{algorithm}
\usepackage{algpseudocode}
\algrenewcommand{\algorithmicrequire}{\textbf{Input:}}
\algrenewcommand{\algorithmicensure}{\textbf{Output:}}
\usepackage{tikz}
\usetikzlibrary{arrows.meta,fit,positioning}
\usepackage[colorlinks=true, linkcolor=blue, citecolor=blue, urlcolor=blue]{hyperref}
\newtheoremstyle{remarkstyle}
    {\topsep}{\topsep}{\normalfont}{}{\bfseries}{}{0.25em}
    {\thmname{#1}\nobreakspace\thmnumber{#2}\thmnote{ (#3)}}
\theoremstyle{remarkstyle}
\newtheorem{remark}{Remark}
\newcommand{\C}{\mathbb{C}}
\newcommand{\R}{\mathbb{R}}
\newcommand{\E}{\mathbb{E}}

\title{Beam-STeP: Generative Site-Specific Beam Prediction with Spatio-Temporal Priors}

\author{Ziyue Wang, Zongyao Zhao, \textit{Member, IEEE}, Zhaolin Wang, \textit{Member, IEEE}, and Yuanwei Liu, \textit{Fellow, IEEE}
	\thanks{Ziyue Wang, Zongyao Zhao, Zhaolin Wang and Yuanwei Liu are with the Department of Electrical and Computer Engineering, The University of Hong Kong, Hong Kong SAR, China (e-mail: ziyue.wang@connect.hku.hk; zongyao@hku.hk; zhaolin.wang@hku.hk; yuanwei@hku.hk).}}

\begin{document}

\maketitle

\begin{abstract}
A generative site-specific beam prediction (Beam-STeP) method is proposed. Spatial priors from site-specific propagation and mobility information are jointly exploited with temporal priors from historical beam observations for beam prediction. To address the beam prediction ambiguity from multimodal spatio-temporal priors, the generative model is adopted, where a history-conditioned diffusion model is proposed to learn the multimodal conditional distribution of future beamforming vectors. Multiple beam candidates are sampled from the distribution using a denoising diffusion implicit model (DDIM). The candidate with the highest received power is then selected for transmission. Simulation results across three urban scenarios demonstrate that: i) Beam-STeP generally achieves higher normalized beamforming gains than the baselines under the same probing budget, particularly as beam prediction progresses; ii) site-specific propagation and mobility priors improve recursive beam prediction performance; and iii) DDIM sampling enables a flexible gain--latency trade-off.

\end{abstract}


\section{Introduction}
\IEEEPARstart{L}{arge}-scale antenna arrays are widely employed in wireless networks to provide directional beamforming gain and support high-data-rate transmission~\cite{Wang2018}. The resulting high-gain beams generally have narrow beamwidths, making the established links sensitive to user mobility and beam misalignment. Even a small displacement of the user equipment (UE) can move the receiver outside the main lobe coverage and cause a substantial degradation in the received signal-to-noise ratio (SNR).

Following initial beam alignment, the serving beam must be updated as the UE moves. Such updates can be obtained through repeated exhaustive beam sweeping or beamformer redesign based on instantaneous channel state information (CSI). However, substantial overhead may be incurred by beam training and CSI acquisition~\cite{Giordani2019}. To reduce this overhead, predictive beam tracking exploits temporal correlations in historical observations to predict suitable beamforming vectors for future time slots.

A straightforward tracking strategy is to probe neighboring beams of the previously selected beam and choose the candidate with the highest received power, as was demonstrated in the field experiments of~\cite{Inoue2015}. Although such local search methods are simple to implement, their performance relies on the assumption that the optimal beam changes gradually and remains close to the previously selected beam. Model-based approaches formulate beam tracking as a recursive state-estimation problem. In particular, extended Kalman filters, unscented Kalman filters, and related Bayesian filters combine a state-transition prior with newly acquired measurements to estimate and predict time-varying channel parameters or beam directions~\cite{Liu2019,Va2016,Larew2019,Jayaprakasam2017}. These methods can achieve efficient tracking with limited measurements, but their accuracy generally depends on the validity of the assumed mobility, channel-evolution, and observation models. Such dependence may limit their robustness when the actual user motion or propagation environment deviates from these assumptions.

Data-driven beam prediction provides an alternative by learning temporal
dependencies between historical observations and future beam states directly
from data. Representative learning-based methods employ recurrent,
attention-based, or other sequence models to infer future beams. For
example, the ODE-LSTM architecture in~\cite{Ma2023a} was developed to
model beam evolution, and~\cite{Ma2021,Zhang2023,Zhao2024} exploited
historical low-frequency channel information or channel estimates to
predict future mmWave beams in different communication scenarios. While learning-based methods have demonstrated promising beam prediction
performance, most existing methods adopt a discriminative formulation
that maps historical observations directly to a single future beamformer.
However, the conditional distribution of future beamforming state in practical
beam tracking is often multimodal due to two common sources of uncertainty.

The first uncertainty source is \textbf{mobility-induced multimodality}. Similar finite beam
histories may correspond to different future trajectories, such as moving
straight, turning left, or turning right at an intersection. These
alternative mobility patterns can lead to substantially different future
beam directions. The second uncertainty source is \textbf{propagation-induced multimodality}.
Limited channel observations provide only an incomplete characterization
of the propagation environment. Consequently, multiple latent propagation
states may be consistent with the same observation history while requiring
different near-optimal beamformers~\cite{Wang2026a}. These two sources of
uncertainty jointly make the mapping from historical observations to future
beamformers inherently one-to-many. Such multimodal uncertainty poses a challenge to the deterministic
prediction strategies which typically produce a single
beam decision from historical observations. For continuous-valued beam predictors trained with the mean-squared-error (MSE) loss, the resulting estimate tends to approximate the conditional mean. When the conditional distribution contains several well-separated beam modes,
the resulting point estimate may average over incompatible complex-weight structures from different modes and fall between physically meaningful high-gain solutions. Such mode averaging can substantially reduce the achievable beamforming gain. 

In contrast, a conditional generative model
learns the distribution of future beamformers and produces multiple
candidate beams through stochastic sampling~\cite{Du2024,Xu2025,VanHuynh2024}.
By generating multiple plausible beam candidates under the same history observations,
the model enables practical beam probing to resolve the uncertainty
associated with the realized mobility and propagation state. Among the works most closely related to generative beam prediction,
\cite{Azarbahram2026} employed a history-conditioned discrete denoising
diffusion probabilistic model (D3PM) to generate future beam indices from
an oversampled codebook using historical probing results and reference
signal received power (RSRP) feedback. The results demonstrated the
potential of generative candidate generation for improving beam prediction
performance.

However, candidate diversity alone does not guarantee that the generated
beams are physically plausible for a particular deployment environment. An
environment-agnostic model may generate candidates pointing toward
inaccessible areas or mobility patterns that are inconsistent with the
local road topology. Since the propagation geometry and mobility
constraints of a fixed BS site remain relatively stable, site-specific
information can be used to restrict the conditional beam distribution to
plausible future states while preserving useful candidate diversity.

Following the principle that ``every site is unique,'' site-specific
beamforming methods have been developed to exploit the geometry and
propagation characteristics of individual deployment environments.
In~\cite{Heng2022}, site-specific probing beams were learned to capture
local propagation features and enable fast prediction of the optimal
narrow beam from limited beam measurements. This idea was further extended
to grid-free MIMO beam alignment in~\cite{Heng2024}, where continuous
transmit and receive beamforming vectors were adopted to overcome the
performance limitation of predefined codebooks. For multi-user MIMO systems, the authors of ~\cite{Zhao2026MU}
exploited SSB-based RSRP fingerprints to infer user spatial
subspaces, enabling user grouping before dedicated CSI
acquisition and low-dimensional CSI refinement for multi-user
beamforming. In~\cite{Abdallah2024}, 
a multi-agent reinforcement learning framework jointly designed active
beam codebooks at the BS and passive reflection codebooks at a
reconfigurable intelligent surface (RIS), while~\cite{Zhang2025c}
developed an intelligent omni-surface (IOS)-specific codebook covering both near-field and far-field
regions for beam alignment. More recently, generative methods for site-specific
beamforming have also been explored. The generative site-specific beamforming (GenSSBF) framework
in~\cite{Wang2026a} highlighted the role of generative models in
capturing site-dependent beam distributions. The Beam Brainstorm solution in~\cite{Zhou2026a} reduced beam-sweeping
overhead by using a small number of RSRP probing measurements to guide
the diffusion-based generation of beams. To address the slow
iterative generation of diffusion-based
methods, the authors of~\cite{Zhou2026b} developed a few-step generator that supports
variable-length probing feedback. In~\cite{Zhao2026}, an
information-maximizing probing codebook was designed with a
conditional flow-matching model to improve the information obtained
from limited RSRP feedback. Furthermore,
the authors of \cite{Jiang2026} extended generative site-specific beamforming to the
multi-user setting using a knowledge-distilled diffusion Transformer that
generates multi-user beamforming weights from low-dimensional RSRP
observations.

These studies demonstrate that site-specific propagation information can
effectively guide generative beam design. Nevertheless, existing site-specific generative methods mainly focus on beamforming or beam alignment based on instantaneous probing to obtain channel observation for current beam generation, which have the following drawbacks: 1) \textit{sufficient instantaneous channel probing is needed for beam generation}, 2) \textit{beam generation requires non-negligible inference time.}

In contrast, beam prediction exploits historical beam sequences to infer future beam states for mobile UEs, which eliminates the need for instantaneous channel observation probing and thus significantly reducing the probing cost.  Beam prediction also allows inference during data transmission, relaxing the time constraints for beam inference. 
Motivated by these advantages, we propose a
site-specific history-conditioned diffusion model that learns the conditional distribution of future beamforming vectors from historical beam observations, where these observations reflect temporal channel evolution and geometry-constrained mobility patterns, providing additional site-specific knowledge for the future beam generation. We then employs a denoising diffusion implicit model (DDIM) sampler to efficiently generate
multiple candidates from the learned distribution during online inference while maintaining a reasonable inference latency.

The main contributions of this paper are summarized as follows:
\begin{itemize}
	\item We propose Beam-STeP, a generative method for site-specific
	beam prediction that jointly exploits spatial and temporal priors. In Beam-STeP,
	spatial priors are implicitly learned during offline training from
	site-specific beam data reflecting the local propagation environment
	and spatial constraints on UE mobility. Temporal priors are
	incorporated through the ordered sequence of historical beamforming
	vectors to capture the recent beam evolution of each UE. Together, these priors guide Beam-STeP toward future beam states that are consistent with both the site-specific environment and the observed beam evolution.
	
	\item We develop a history-conditioned diffusion model for Beam-STeP
	to address the multimodal nature of future beam evolution, where
	the same beam history may correspond to multiple plausible future
	beam states. Specifically, a denoising diffusion probabilistic model
	(DDPM) is conditioned on the historical beamforming-vector sequence
	to learn the conditional distribution of future beams. During online inference,
	DDIM sampling generates multiple beamforming vector candidates
	using substantially fewer denoising steps than conventional DDPM
	sampling. The candidates are then probed, and the one with the highest
	measured RSRP is selected for data transmission.

	\item We conduct extensive simulations across three urban scenarios, incorporating a road-aware mobility model to capture geometry-constrained UE motion, and evaluate Beam-STeP under different probing budgets and site-specific configurations. The results demonstrate improved prediction performance over the deterministic ODE-LSTM baseline and highlight the gain benefits of continuous beam generation and site-specific training.
\end{itemize}

The rest of this paper is organized as follows. Section II presents the system model and optimization objective. Section III describes the proposed Beam-STeP generative method. Section IV reports numerical results, and Section V concludes the paper.

\emph{Notations:} Scalars, vectors, and matrices are denoted by regular italic, bold lowercase, and bold uppercase letters, respectively. The real and complex fields are denoted by $\R$ and $\C$. The transpose and Hermitian transpose are represented by $(\cdot)^T$ and $(\cdot)^H$, and $j=\sqrt{-1}$ is the imaginary unit. The magnitude and Euclidean norm are denoted by $|\cdot|$ and $\|\cdot\|_2$, and the phase is denoted by $\angle(\cdot)$. The matrix $\bm I$ denotes the identity matrix of appropriate dimension. The distributions $\mathcal N(\bm\mu,\bm\Sigma)$ and $\mathcal{CN}(\bm\mu,\bm\Sigma)$ denote real and circularly symmetric complex Gaussian distributions, respectively, and expectation is denoted by $\E[\cdot]$.

\section{System Model and Problem Formulation}
\subsection{Site-Specific Downlink Channel}
We consider a narrowband downlink communication system in which one BS equipped with an $N_T$-element uniform linear array (ULA) serves single-antenna UEs over frequency-flat fading channels. The fixed deployment site denoted by $\mathcal S$ is characterized by the BS placement and array orientation, the surrounding geometry and material properties, and the road topology that constrains UE mobility. These persistent site features determine both the site-specific propagation statistics and the feasible mobility patterns. 

At time slot $t$, the downlink channel vector in site $\mathcal S$ with $L_t$ propagation paths is given by
\begin{equation}
    \bm h_t = \sum_{\ell=1}^{L_t} \sqrt{p_{t,\ell}}
    e^{j\varphi_{t,\ell}}\bm a(\theta_{t,\ell}) \in \C^{N_T\times 1},
    \label{eq:channel}
\end{equation}
where $p_{t,\ell}$, $\varphi_{t,\ell}$, $\theta_{t,\ell}$ denote the power, phase, and angle of departure of path $\ell$, respectively. The normalized array response at departure angle $\theta$ is given by
\begin{equation}
    \bm a(\theta)=\frac{1}{\sqrt{N_T}}
    \left[1,e^{j2\pi\frac{d}{\lambda}\sin\theta},\ldots,
    e^{j2\pi\frac{d}{\lambda}(N_T-1)\sin\theta}\right]^{T}.
    \label{eq:steering}
\end{equation}
Here, $\lambda$ is the carrier wavelength and $d$ represents the inter-antenna spacing. The time dependence of the path parameters captures the channel evolution induced by UE motion, while their distribution is governed by $\mathcal S$.

\begin{figure}[t]
	\centering
	\includegraphics[width=0.92\columnwidth]{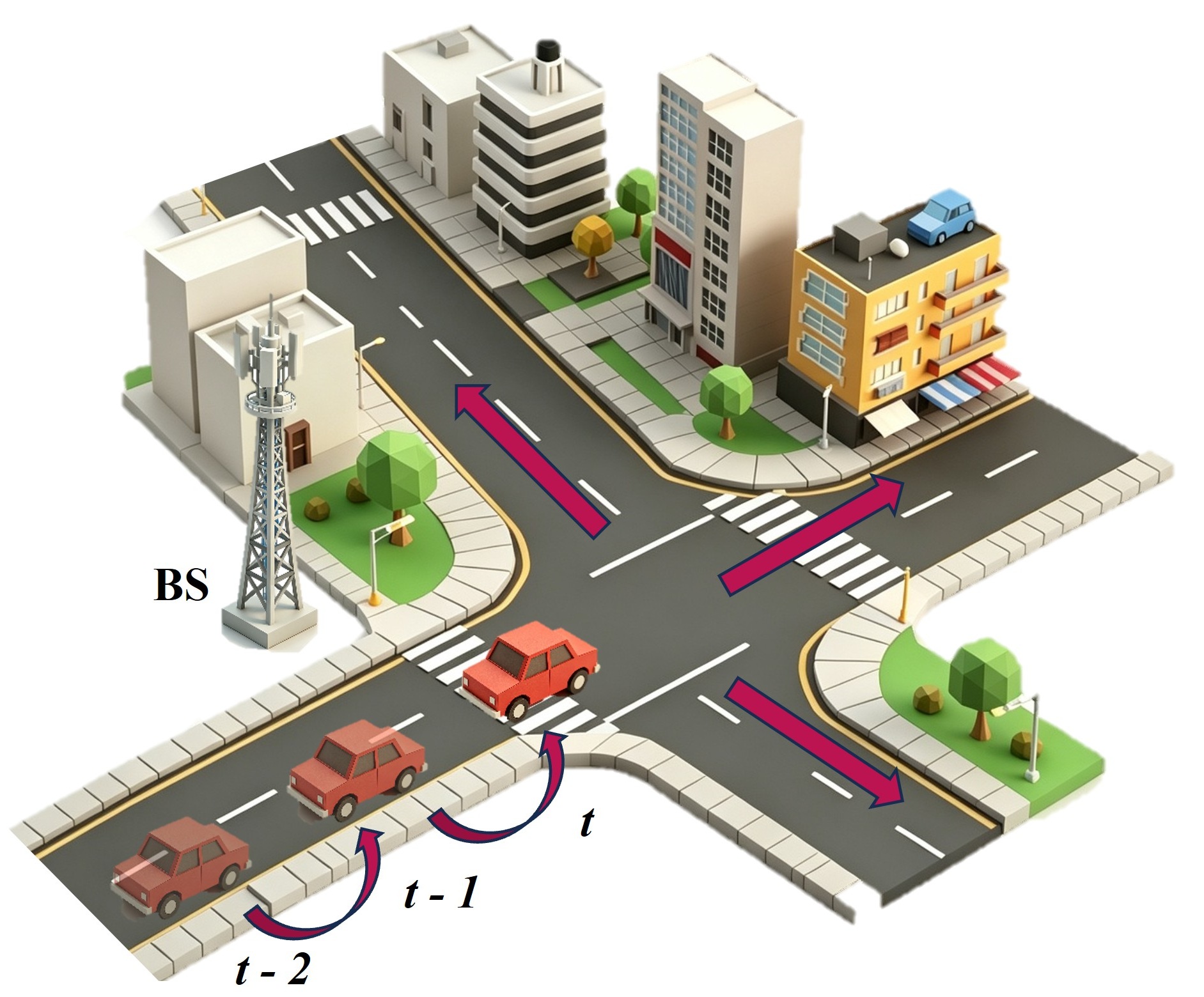}
	\caption{Beam prediction for next time slot under certain site-specific environment.}
	\label{fig:system}
\end{figure}

We consider a analog beamforming architecture, in which a phase-shifter network controls the transmit phase at each antenna element. As a consequence, the transmit beamforming vector at time slot $t$ can be written as
\begin{equation}
    \bm w_t=\frac{1}{\sqrt{N_T}}
    \left[e^{j\phi_{t,0}},e^{j\phi_{t,1}},\ldots,
    e^{j\phi_{t,N_T-1}}\right]^T\in\C^{N_T\times 1},
    \label{eq:analog_beam}
\end{equation}
where $\phi_{t,n}$ is the phase of the $n$-th antenna element. Accordingly, every feasible analog beam should satisfy the constant-modulus constraint imposed by hardware which as follows:
\begin{equation}
    \mathcal W_{\rm C}=\left\{\bm w\in\C^{N_T}: |w_n|=\frac{1}{\sqrt{N_T}},\ n=0,\ldots,N_T-1\right\},
    \label{eq:cm_set}
\end{equation}
where $w_n$ is the $n$-th entry of $\bm w$. Let $s_t$ denote a unit-power transmitted data symbol satisfying $\E[|s_t|^2]=1$. Therefore, the complex baseband signal received by the UE can be written as
\begin{equation}
    y_t = \sqrt{P_{\rm T}}\bm h_t^{H}\bm w_t s_t+n_t,
    \label{eq:received}
\end{equation}
where $P_{\rm T}$ is the BS transmit power and $n_t\sim\mathcal{CN}(0,\sigma^2)$ is circularly symmetric complex Gaussian noise with variance $\sigma^2$ at time slot $t$. The corresponding receive SNR is
\begin{equation}
    \gamma_t(\bm w)=\frac{P_{\rm T}}{\sigma^2}
    |\bm h_t^H\bm w|^2.
    \label{eq:snr}
\end{equation}
Since $P_{\rm T}/\sigma^2$ is independent of the beamforming vector, maximizing the received SNR is equivalent to maximizing the following beamforming gain:
\begin{equation}
    g_t(\bm w)=|\bm h_t^H\bm w|^2.
    \label{eq:gain}
\end{equation}
When the instantaneous channel $\bm h_t$ is perfectly known, the optimal beamforming vector that maximizes \eqref{eq:gain} is given by the following maximum ratio transmission (MRT) solution:
\begin{equation}
    \bm w_t^\star
    =\underset{\bm w\in\mathcal W_{\rm C}}{\arg\max}\,
    g_t(\bm w)
    =\frac{1}{\sqrt{N_T}}\exp\!\left(j\angle \bm h_t\right),
    \label{eq:phase_optimal}
\end{equation}
which serves as the performance upper bound of the system. In practice, acquiring
accurate instantaneous $\bm h_t$ incurs channel-estimation overhead, while UE
mobility can render the acquired CSI outdated before it is used for
beamforming. This motivates predicting future beam candidates
from historical observations and selecting the best candidate through
limited probing, without explicitly estimating the full instantaneous
channel.

\begin{figure*}[t]
	\centering
	\includegraphics[width=\textwidth]{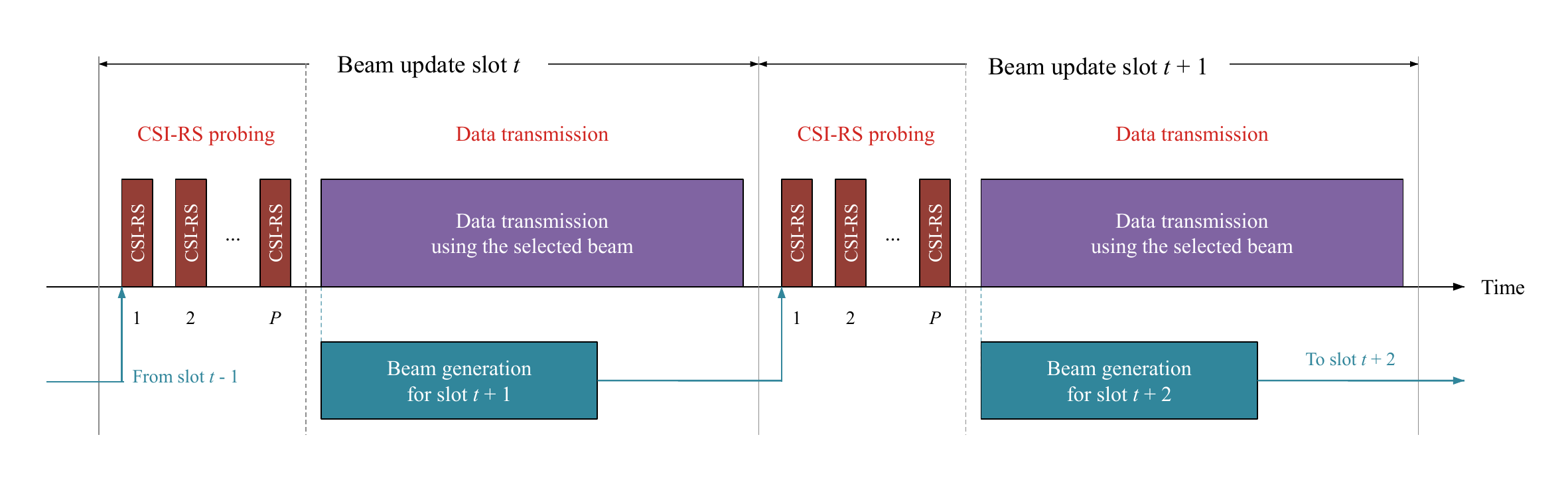}
	\caption{Illustration of the beam prediction.}
	\label{fig:beam_slot_timeline}
\end{figure*}

\subsection{Beam Prediction and Problem Formulation}
We consider a beam prediction setting in which the BS determines a set of $P$ feasible beamforming candidates for the next time slot. At the beginning of each slot, the candidates generated in the preceding slot are evaluated through CSI-RS probing, and the candidate with the highest measured RSRP is selected for data transmission. During data transmission, the BS uses the updated beam history to generate candidates for the following slot. Within each probing round, the effective channel is assumed to remain quasi-static, such that all $P$ candidate beams are evaluated under the same channel realization $\bm h_t$. In contrast, the channel is
assumed to evolve across consecutive slots due to UE mobility. Fig.~\ref{fig:beam_slot_timeline} illustrates this operation, where candidate generation is assumed to finish within the current data-transmission period.

Let $\bar{\bm w}_\tau\in\mathcal W_{\rm C}$ denote the historical beamforming vector selected and stored by the beam-tracking procedure at slot $\tau$. Given a observation window length $T_{\rm h}$, the temporal prior for predicting the beams at time slot $t+1$ is given by
\begin{equation}
	\mathcal H_t=
	\left[
	\bar{\bm w}_{t-T_{\rm h}+1},
	\ldots,
	\bar{\bm w}_t
	\right].
	\label{eq:beam_history}
\end{equation}
The future channel $\bm h_{t+1}$ and its optimal beam
$\bm w_{t+1}^\star$ are unavailable when the prediction is made.
The BS therefore determines future beam candidates based on
the available history observations $\mathcal H_t$ and site-specific prior information.
Let $p_{\mathcal S}(\cdot\mid\mathcal H_t)$ denote a conditional
distribution over feasible beamforming vectors to be designed for
site $\mathcal S$. For a probing budget of $P$, the BS independently
samples $P$ candidates from this distribution to form
\begin{equation}
    \widehat{\mathcal W}_{t+1}
    =\left\{\widehat{\bm w}_{t+1}^{(m)}\right\}_{m=1}^{P},
    \qquad
    \widehat{\bm w}_{t+1}^{(m)}\sim
    p_{\mathcal S}(\bm w_{t+1}\mid\mathcal H_t),
    \label{eq:beam_candidate_set}
\end{equation}
where every candidate satisfies the hardware constraint $\widehat{\bm w}_{t+1}^{(m)}\in\mathcal W_{\rm C}$. Then, the candidate with largest gain will be finalized as
\begin{equation}
    m^\star=\arg\max_{1\leq m\leq P}
    g_{t+1}\!\left(\widehat{\bm w}_{t+1}^{(m)}\right),
    \qquad
    \widehat{\bm w}_{t+1}
    =\widehat{\bm w}_{t+1}^{(m^\star)}.
    \label{eq:selection}
\end{equation}
To evaluate candidate quality across different channel conditions, we normalize the beamforming gain of the best candidate by the optimal constant-modulus beamforming gain for each channel realization. This measures relative beamforming performance and prevents strong-channel samples from dominating the objective solely because of their larger channel magnitudes. The site-specific beam prediction problem is therefore formulated as an expected best-of-$P$ normalized beamforming-gain maximization problem:
\begin{subequations}
\label{eq:problem}
\begin{align}
    \underset{p_{\mathcal S}}{\operatorname{max}}\quad
    &J_P(p_{\mathcal S})
    \triangleq
    \E
    \!\left[
    \max_{1\leq m\leq P}
    \frac{g_{t+1}\!\left(\widehat{\bm w}_{t+1}^{(m)}\right)}
    {g_{t+1}\!\left(\bm w_{t+1}^\star\right)}
    \right]
    \label{eq:problem_objective}\\
    \text{s.t.}\quad
    &\widehat{\bm w}_{t+1}^{(m)}\sim
    p_{\mathcal S}(\bm w_{t+1}\mid\mathcal H_t),
    \quad m=1,\ldots,P,
    \label{eq:problem_distribution}\\
    &\widehat{\bm w}_{t+1}^{(m)}\in\mathcal W_{\rm C},
    \quad m=1,\ldots,P,
    \label{eq:problem_feasibility}
\end{align}
\end{subequations}
where the expectation in \eqref{eq:problem_objective} is over the site-specific history--channel distribution and the randomness of candidate generation under $p_{\mathcal S}$. Constraint \eqref{eq:problem_distribution} specifies sampling conditioned on the available history observations, and \eqref{eq:problem_feasibility} imposes the constant-modulus analog-beamforming requirement.

\section{Proposed Beam-STeP Method}
In this section, we present Beam-STeP, a history-conditioned diffusion method to the site-specific beam prediction problem. During the offline training process, site-specific historical and future beamforming vectors are used to train a conditional noise predictor under a Gaussian diffusion process. During online inference, a DDIM sampler is deployed to generate $P$ beam candidates from independent Gaussian initializations based on the observed beam history. The generated samples are first reconstructed as feasible analog beamformers, after which the candidates are probed and the beam with the highest RSRP is selected for data transmission. The term Beam-STeP reflects the joint exploitation of \underline{S}patial priors from the deployment site and \underline{T}emporal priors from historical beam evolution, \underline{e}nabling efficient site-specific beam \underline{P}rediction. The details of the algorithm are presented in the following subsections. Section~\ref{sec:angular_representation} introduces the angular-domain beam representation. Section~\ref{sec:diffusion_formulation} presents the conditional diffusion process. Section~\ref{sec:network_architecture} describes the proposed network architecture. Section~\ref{sec:training_deployment} details the offline training and online deployment procedures.

\subsection{Angular-Domain Beam Representation}
\label{sec:angular_representation}
Before a beamforming vector is processed by the network, it is transformed into the angular domain. In the spatial domain, the beam phase information
is distributed across the antenna elements, and the beam direction is
not explicit in the individual vector elements. In the angular domain, each vector element represents the projection of the beamforming vector onto a spatial basis vector, and its squared magnitude measures the
corresponding projection energy. This representation makes the directional structure of the beamforming
vector more explicit. Specifically, we apply a unitary discrete Fourier transform (DFT) and retain all coefficients, making the operation an invertible representation change. Let $\bm F$ denote the unitary
DFT matrix, which is given by
\begin{equation}
    [\bm F]_{q,n}=\frac{1}{\sqrt{N_T}}
    \exp\left(-j\frac{2\pi qn}{N_T}\right),
    \quad q,n=0,\ldots,N_T-1,
    \label{eq:dft_matrix}
\end{equation}
where $q$ and $n$ are the angular-bin and antenna-element indices, respectively. Since $\bm F^H\bm F=\bm I$, the spatial and angular-domain beams are related by
\begin{equation}
    \bm w_t^{\rm ang}=\bm F\bm w_t,
    \qquad
    \bm w_t=\bm F^H\bm w_t^{\rm ang}.
    \label{eq:angular_transform}
\end{equation}
For a complex angular vector $\bm w^{\rm ang}\in\C^{N_T}$, define the real-valued representation operator
\begin{equation}
    \rho(\bm w^{\rm ang})=
    \begin{bmatrix}\Re\{\bm w^{\rm ang}\}^T &
    \Im\{\bm w^{\rm ang}\}^T\end{bmatrix}^T\in\R^{2N_T\times 1},
    \label{eq:ri_operator}
\end{equation}
where $\Re\{\cdot\}$ and $\Im\{\cdot\}$ denote the real part and imaginary part, respectively. Therefore, the input data of the model can be written as
\begin{equation}
    \bm X_t=\left[
    \rho(\bm F\bar{\bm w}_{t-T_{\rm h}+1}),\ldots,
    \rho(\bm F\bar{\bm w}_{t})
    \right]^T\in\R^{T_{\rm h}\times 2N_T}.
    \label{eq:history}
\end{equation}
Accordingly, the training target can be represented as
\begin{equation}
    \bm x_{0,t+1}=\rho(\bm F\bm w_{t+1}^{\star}),
    \label{eq:angular_target}
\end{equation}
where the diffusion-level index $0$ indicates an uncorrupted sample. Conversely, any real-valued angular representation $\bm x\in\R^{2N_T}$ is reconstructed in the antenna domain as
\begin{equation}
    \widetilde{\bm w}=\bm F^H\rho^{-1}(\bm x).
    \label{eq:angular_reconstruction}
\end{equation}
Since $\widetilde{\bm w}$ is not necessarily phase-only, it is projected onto the feasible analog-beam set as
\begin{equation}
    \mathcal P_{\rm CM}(\widetilde{\bm w})
    =\frac{1}{\sqrt{N_T}}\exp\!\left(j\angle\widetilde{\bm w}\right).
    \label{eq:cm_projection}
\end{equation}
Here, $\rho^{-1}$ reconstructs a complex angular vector from its real and imaginary parts, and $\mathcal P_{\rm CM}$ denotes elementwise projection onto $\mathcal W_{\rm C}$. The conditional diffusion model subsequently generates angular-domain samples that are converted to physical beam candidates through \eqref{eq:angular_reconstruction} and \eqref{eq:cm_projection}.

\subsection{History-Conditioned Diffusion Training and DDIM Sampling}
\label{sec:diffusion_formulation}
To capture the multimodal nature of future beam evolution,
we employ a diffusion model conditioned on the historical
beamforming-vector sequence.
During training, a forward Gaussian process corrupts the clean
angular-domain beam, and a conditional denoiser learns to predict the
injected noise at arbitrary noise levels~\cite{Ho2020}. During online generation, a DDIM sampler is deployed to transform independent Gaussian
initializations to multiple plausible future-beam candidates conditioned on
the beam history observations~\cite{Song2021}.
\begin{figure*}[t]
	\centering
	\includegraphics[width=\textwidth]{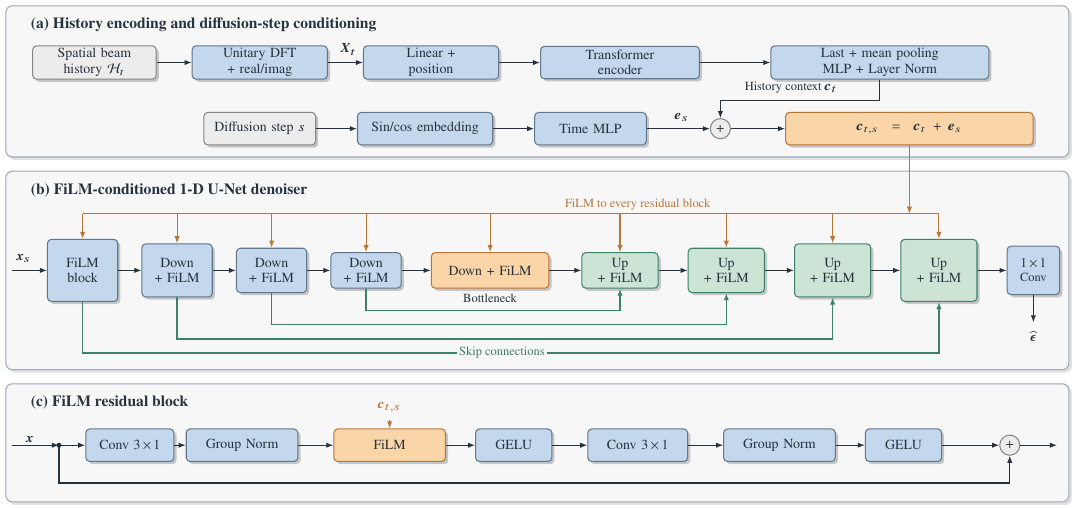}
	\caption{Architecture of the noise-prediction network in the proposed Beam-STeP method.}
	\label{fig:architecture}
\end{figure*}
Let $\bm\Theta_{\mathcal S}$ denote the trainable neural network parameters
of the site-specific diffusion model. The model
learns the site-specific conditional distribution of the next angular beam as
\begin{equation}
    p_{\bm\Theta_{\mathcal S}}(\bm x_{0,t+1}\mid\bm X_t).
    \label{eq:conditional_distribution}
\end{equation}

Let $\bm x_0=\bm x_{0,t+1}=\rho(\bm F\bm w_{t+1}^{\star})\in\R^{2N_T}$.
The offline training dataset at site $\mathcal S$ is denoted by
$\mathcal D_{\mathcal S}$ and consists of angular-domain history--target
pairs $(\bm X_t,\bm x_0)$.
A forward diffusion process corrupts the beam vector over $N_{\mathrm{diff}}$ steps, where $N_{\mathrm{diff}}$ is the total number of diffusion steps and $s\in\{1,\ldots,N_{\mathrm{diff}}\}$ is the diffusion-step index. Given a variance schedule $\{\beta_s\}_{s=1}^{N_{\mathrm{diff}}}$ and $\alpha_s=1-\beta_s$, each forward transition is
\begin{equation}
    q(\bm x_s\mid\bm x_{s-1})=
    \mathcal N\!\left(
    \bm x_s;\sqrt{\alpha_s}\bm x_{s-1},\beta_s\bm I
    \right).
    \label{eq:forward_transition}
\end{equation}
The complete forward Markov chain is therefore factorized as
\begin{equation}
    q(\bm x_{1:N_{\mathrm{diff}}}\mid\bm x_0)
    =\prod_{s=1}^{N_{\mathrm{diff}}}q(\bm x_s\mid\bm x_{s-1}).
    \label{eq:forward_chain}
\end{equation}
Defining $\bar\alpha_s=\prod_{i=1}^{s}\alpha_i$, a noisy sample at any diffusion step can be obtained directly from the clean target as
\begin{equation}
    \bm x_s=\sqrt{\bar\alpha_s}\bm x_0+
    \sqrt{1-\bar\alpha_s}\bm\epsilon,\qquad
    \bm\epsilon\sim\mathcal N(\bm 0,\bm I),
    \label{eq:forward_diffusion}
\end{equation}
which avoids simulating all preceding forward transitions during training.
The denoiser
$\bm\epsilon_{\bm{\Theta}_{\mathcal S}}(\bm x_s,s,\bm X_t)$ predicts
the injected noise conditioned on the beam history observations and is optimized using
\begin{equation}
\begin{aligned}
    \mathcal L_{\rm diff}(\bm{\Theta}_{\mathcal S})
    =\E\Big[&\left\|\bm\epsilon-
    \bm\epsilon_{\bm{\Theta}_{\mathcal S}}
    (\bm x_s,s,\bm X_t)\right\|_2^2\Big],\\[-1mm]
    &(\bm X_t,\bm x_0)\sim\mathcal D_{\mathcal S},\quad
    s\sim\mathcal U\{1,\ldots,N_{\mathrm{diff}}\},\\[-1mm]
    &\bm\epsilon\sim\mathcal N(\bm0,\bm I),
\end{aligned}
    \label{eq:diffusion_loss}
\end{equation}
where $\mathcal U\{1,\ldots,N_{\mathrm{diff}}\}$ denotes the discrete uniform distribution over diffusion steps. The noise-prediction loss $\mathcal L_{\mathrm{diff}}$ in \eqref{eq:diffusion_loss} provides a surrogate objective for training, while the best-of-$P$ normalized beamforming gain in \eqref{eq:problem_objective} serves as the performance criterion after probing and selection. 

During inference, DDIM sampling uses the trained denoiser to generate beam candidates with fewer reverse steps. It preserves the forward-process marginals while replacing the stochastic adjacent-step reverse chain with an implicit trajectory over a subset of the training diffusion levels. Specifically, denote a shortened schedule containing $K$ reverse updates as
\begin{equation}
\begin{aligned}
    \mathcal T_K&=\{\tau_K,\tau_{K-1},\ldots,\tau_1\},\\
    N_{\mathrm{diff}}=\tau_K&>\tau_{K-1}>\cdots>\tau_1>\tau_0=0,
\end{aligned}
    \label{eq:ddim_timestep_sequence}
\end{equation}
where $K\ll N_{\mathrm{diff}}$
and $\bar\alpha_{\tau_0}=1$. Starting from
$\bm x_{\tau_K}=\bm x_{N_{\mathrm{diff}}}\sim\mathcal N(\bm0,\bm I)$, the predicted noise at
the $i$-th update is
\begin{equation}
    \widehat{\bm\epsilon}_i=
    \bm\epsilon_{\bm\Theta_{\mathcal S}}
    (\bm x_{\tau_i},\tau_i,\bm X_t).
    \label{eq:ddim_predicted_noise}
\end{equation}
The corresponding clean-beam estimate is then obtained as
\begin{equation}
    \widehat{\bm x}_{0\mid\tau_i}
    =\frac{\bm x_{\tau_i}
    -\sqrt{1-\bar\alpha_{\tau_i}}\widehat{\bm\epsilon}_i}
    {\sqrt{\bar\alpha_{\tau_i}}}.
    \label{eq:ddim_clean_estimate}
\end{equation}
The stochasticity of the DDIM update is controlled by
\begin{equation}
    \sigma_i=\eta\sqrt{
    \frac{1-\bar\alpha_{\tau_{i-1}}}
         {1-\bar\alpha_{\tau_i}}
    \left(1-
    \frac{\bar\alpha_{\tau_i}}
         {\bar\alpha_{\tau_{i-1}}}\right)},
    \label{eq:ddim_sigma}
\end{equation}
and the update from $\tau_i$ directly to $\tau_{i-1}$ is
\begin{equation}
\begin{aligned}
    \bm x_{\tau_{i-1}}
    ={}&\sqrt{\bar\alpha_{\tau_{i-1}}}
    \widehat{\bm x}_{0\mid\tau_i}
    +\sqrt{1-\bar\alpha_{\tau_{i-1}}-\sigma_i^2}
    \widehat{\bm\epsilon}_i\\
    &+\sigma_i\bm z_i,
    \qquad \bm z_i\sim\mathcal N(\bm0,\bm I),
\end{aligned}
    \label{eq:reverse}
\end{equation}
for $i=K,K-1,\ldots,1$. The parameter $\eta\in[0,1]$ controls the
additional randomness introduced along the reverse trajectory. In the proposed implementation, we set $\eta=0.3$.
Candidate generation therefore involves randomness from
both independent Gaussian initializations and the
Gaussian noise injected during the reverse updates.

\subsection{Network Architecture}
\label{sec:network_architecture}
The noise predictor $\bm\epsilon_{\bm\Theta_{\mathcal S}}(\bm x_s,s,\bm X_t)$ comprises a Transformer-based history encoder, a diffusion-step encoder, and a feature-wise linear modulation (FiLM)-conditioned one-dimensional U-Net denoiser.

\subsubsection{History Encoder}

The history encoder first projects each angular real/imaginary vector into a $D$-dimensional token, where $D$ is the Transformer embedding dimension, and adds a learned positional embedding. A pre-normalized Transformer produces token states $\{\bm u_1,\ldots,\bm u_{T_{\rm h}}\}$. To preserve both the most recent state and the trajectory-wide tendency, context pooling uses
\begin{equation}
    \bm c_t=\operatorname{LN}\!\left(
    \operatorname{MLP}\!\left[
    \bm u_{T_{\rm h}}\,\Vert\,
    \frac{1}{T_{\rm h}}\sum_{\tau=1}^{T_{\rm h}}\bm u_\tau
    \right]\right).
    \label{eq:history_context}
\end{equation}
Here, $\operatorname{MLP}(\cdot)$ and $\operatorname{LN}(\cdot)$ denote a multilayer perceptron and layer normalization, respectively, and $\Vert$ denotes vector concatenation. The last token preserves the most recent beam state, whereas mean pooling summarizes the overall evolution across the observation window. An MLP is applied to the concatenated features to jointly encode
the most recent beam state and the overall history observations into a unified context representation.

\subsubsection{Diffusion-Step Encoder}
For this module, a sinusoidal embedding of diffusion step $s$ is simply processed by a time MLP to produce $\bm e_s$, which identifies the current noise level.

\subsubsection{FiLM Conditioning}
The diffusion-step embedding is fused with the history context generated by the history encoder to form
\begin{equation}
    \bm c_{t,s}=\bm c_t+\bm e_s.
    \label{eq:condition}
\end{equation}
The resulting condition informs the denoiser of both the site-specific temporal context and the amount of corruption to be removed at step $s$. The condition is injected into every U-Net residual convolution block through FiLM~\cite{Perez2018}. After the first normalization, it modulates the feature map $\bm z$ as
\begin{equation}
\operatorname{FiLM}_{\ell}(\bm z;\bm c_{t,s})
=
\bm z\odot
\left(1+\bm\zeta_{\ell}(\bm c_{t,s})\right)
+\bm\xi_{\ell}(\bm c_{t,s}),
    \label{eq:film}
\end{equation}
where $\bm\zeta_{\ell}(\cdot)$ and $\bm\xi_{\ell}(\cdot)$ are the
learned channel-wise scaling and shifting projections, respectively,
for the $\ell$-th residual block.

\subsubsection{U-Net}
The FiLM-conditioned U-Net receives the noisy angular-domain sample
$\bm x_s\in\mathbb{R}^{2N_T \times 1}$, which is reshaped into a two-channel
one-dimensional feature map of size $2\times N_T$ corresponding to the
real and imaginary parts. Each residual block contains two
length-preserving one-dimensional convolutions,
group normalization, Gaussian Error Linear Unit (GELU) activations, and a residual projection. The
history--time condition $\bm c_{t,s}$ is injected into these blocks through
the FiLM operation introduced above.

The encoder contains four stride-two stages, reducing the angular-bin
length while increasing the hidden
channel width. By progressively enlarging the receptive field, the encoder captures both localized angular peaks and broader cross-bin dependencies in the noisy beam pattern. The decoder then restores the angular resolution using transposed convolutions and U-Net skip connections. The skip connections preserve fine-grained angular information during reconstruction, which is important for accurately locating high-gain beam directions. Finally, a $1\times1$ convolution maps the features back to two channels,
which are flattened to obtain the predicted noise
$\widehat{\bm\epsilon}\in\mathbb{R}^{2N_T \times 1}$. The overall architecture
is summarized in Fig.~\ref{fig:architecture}.

\subsection{Offline Training and Online Deployment}
\label{sec:training_deployment}
\subsubsection{Offline Training}
Offline training uses angular-domain history--target pairs $(\bm X_t,\bm x_0)$ from $\mathcal D_{\mathcal S}$. In Beam-STeP, site-specific propagation and mobility information is implicitly encoded in the model parameters through training on $\mathcal D_{\mathcal S}$, without requiring an explicit map input during online inference. For each mini-batch, a diffusion step $s$ and Gaussian noise $\bm\epsilon$ are sampled, and the noisy target $\bm x_s$ is obtained directly from \eqref{eq:forward_diffusion}. The network predicts the injected noise from $(\bm x_s,s,\bm X_t)$, and $\bm\Theta_{\mathcal S}$ is updated by minimizing \eqref{eq:diffusion_loss}. This exposes the denoiser to all noise levels without traversing the complete forward chain for every training pair.
\begin{figure}[t]
    \centering
    \subfloat[\label{fig:eyebird}]{%
        \includegraphics[height=0.38\columnwidth]{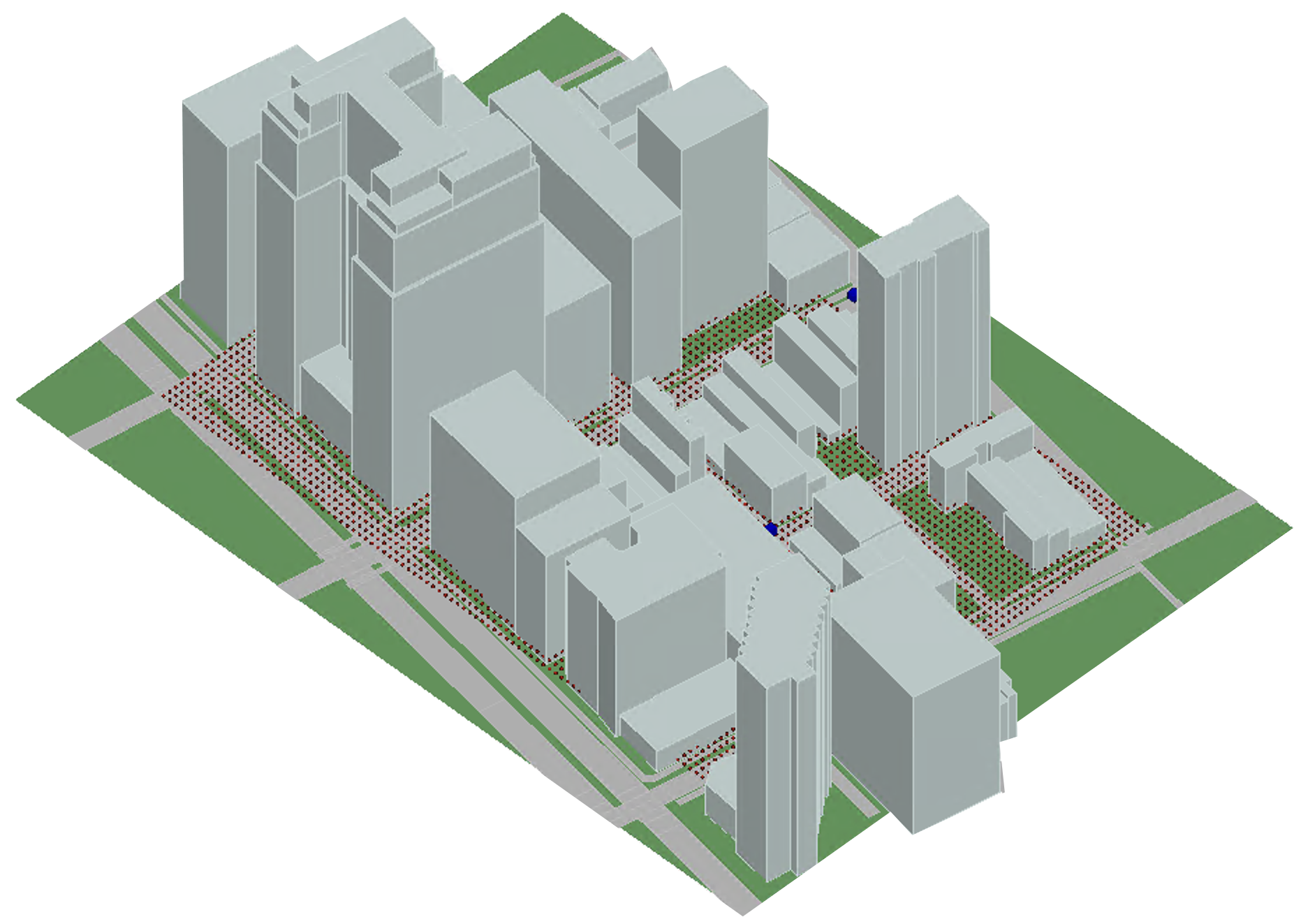}}%
    \hfill
    \subfloat[\label{fig:satellite}]{%
        \includegraphics[height=0.38\columnwidth]{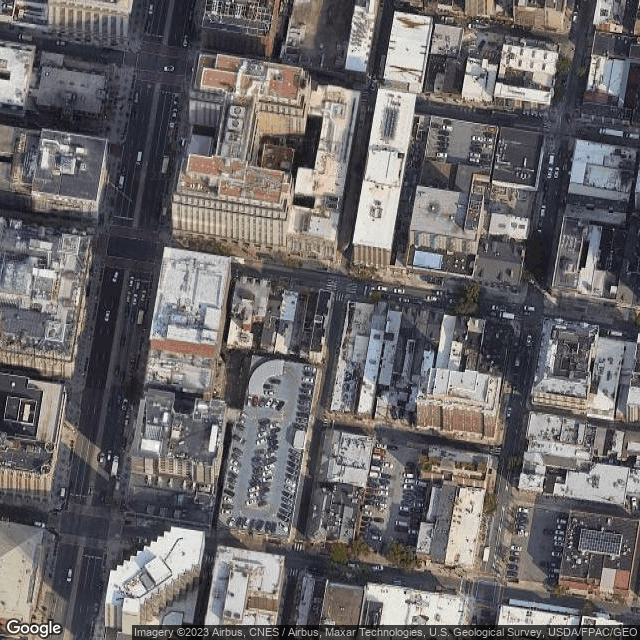}}
    \caption{The city\_5\_philadelphia\_28 scenario: (a) the bird's-eye view and (b) the satellite view.}
    \label{fig:philadelphia_site}
\end{figure}
\subsubsection{Online Inference}
Given the current history observations $\bm X_t$, each candidate is independently generated as
\begin{equation}
    \widehat{\bm x}_{0,t+1}^{(m)}\sim
    p_{\bm\Theta_{\mathcal S}}(\cdot\mid\bm X_t),
    \quad m=1,\ldots,P.
    \label{eq:candidate_sampling}
\end{equation}
In implementation, each candidate begins from an independent $\bm x_{N_{\mathrm{diff}}}^{(m)}\sim\mathcal N(\bm0,\bm I)$, and repeated application of \eqref{eq:reverse} produces $\widehat{\bm x}_{0,t+1}^{(m)}$. Its feasible physical beam is then obtained as $\widehat{\bm w}_{t+1}^{(m)}=\mathcal P_{\rm CM}\!\left(\bm F^H\rho^{-1}(\widehat{\bm x}_{0,t+1}^{(m)})\right)$ using \eqref{eq:angular_reconstruction} and \eqref{eq:cm_projection}. Repeating this procedure for $m=1,\ldots,P$ yields the candidate set in \eqref{eq:beam_candidate_set}. This generation process runs concurrently with data transmission in slot $t$. At the beginning of slot $t+1$, the BS probes the prepared candidates, and the UE reports the highest RSRP candidate according to \eqref{eq:selection} for data transmission in that slot.

The offline training and online inference of the proposed Beam-STeP are summarized in Algorithm~\ref{alg:offline_training} and
Algorithm~\ref{alg:online_inference}, respectively. To predict the following step, the beam selected after probing is transformed into its angular-domain representation and appended to the history, while the oldest history token is removed. The updated history observations is then used as the condition for the next prediction.

\begin{algorithm}[t]
    \caption{Offline Training of Beam-STeP}
    \label{alg:offline_training}
    \small
    \begin{algorithmic}[1]
        \Require Site-specific dataset $\mathcal D_{\mathcal S}$; training schedule $\{\beta_s\}_{s=1}^{N_{\mathrm{diff}}}$
        \Ensure Trained parameters $\bm\Theta_{\mathcal S}$
        \State Initialize the network parameters $\bm\Theta_{\mathcal S}$
        \Repeat
            \State Sample a mini-batch of $(\bm X_t,\bm x_0)$ from $\mathcal D_{\mathcal S}$
            \State For each pair, independently sample $s\sim\mathcal U\{1,\ldots,N_{\mathrm{diff}}\}$ and $\bm\epsilon\sim\mathcal N(\bm0,\bm I)$
            \State Construct the noisy target $\bm x_s$ using \eqref{eq:forward_diffusion}
            \State Predict $\widehat{\bm\epsilon}=\bm\epsilon_{\bm\Theta_{\mathcal S}}(\bm x_s,s,\bm X_t)$
            \State Update $\bm\Theta_{\mathcal S}$ by minimizing the mini-batch noise-prediction loss in \eqref{eq:diffusion_loss}
        \Until{the training stopping criterion is met}
        \State Retain the parameters with the lowest validation loss
    \end{algorithmic}
\end{algorithm}

\begin{algorithm}[t]
    \caption{Online Inference of Beam-STeP}
    \label{alg:online_inference}
    \small
    \begin{algorithmic}[1]
        \Require Trained parameters $\bm\Theta_{\mathcal S}$; initial history observations $\mathcal H_t$; probing budget $P$; DDIM schedule $\mathcal T_K$ and stochasticity $\eta$
        \Ensure Selected transmission beams $\{\widehat{\bm w}_{t+1}\}$
        \State Form $\bm X_t$ from the initial history observations $\mathcal H_t$ using \eqref{eq:history}
        \While{the UE is being served}
            \State Encode $\bm X_t$ once to obtain $\bm c_t$ using \eqref{eq:history_context}
            \State Independently initialize $\bm x_{\tau_K}^{(m)}\sim\mathcal N(\bm0,\bm I)$ for $m=1,\ldots,P$
            \For{$i=K,K-1,\ldots,1$}
                \State Predict noise for all $P$ candidates in parallel using \eqref{eq:ddim_predicted_noise} and cached $\bm c_t$
                \State Apply the DDIM update in \eqref{eq:ddim_clean_estimate}--\eqref{eq:reverse} with independent noise for each candidate
            \EndFor
            \State Convert $\{\bm x_{\tau_0}^{(m)}\}_{m=1}^{P}$ to feasible beam candidates using \eqref{eq:angular_reconstruction} and \eqref{eq:cm_projection}
            \State At slot $t+1$, probe the candidates and use the highest-RSRP beam $\widehat{\bm w}_{t+1}$ for transmission, as in \eqref{eq:selection}
            \State Form $\bm X_{t+1}$ by dropping the oldest row of $\bm X_t$ and appending $\rho(\bm F\widehat{\bm w}_{t+1})^T$
            \State $t\gets t+1$
        \EndWhile
    \end{algorithmic}
\end{algorithm}

\section{Numerical Results}
In this section, we evaluate the site-specific performance of Beam-STeP in three scenarios. A road-aware trajectory construction method is adopted to simulate the real UE mobility pattern. The numerical results are then presented and analyzed.
\subsection{Simulation Setup}
The experiments are conducted using the DeepMIMO ray-tracing dataset~\cite{deepmimo}, including the following three scenarios: \textit{\textbf{city\_5\_philadelphia\_28}}, \textit{\textbf{city\_17\_seattle\_28}}, and \textit{\textbf{city\_2\_chicago\_28}}. Fig.~\ref{fig:philadelphia_site} shows the bird's-eye view of the ray-tracing scene and the corresponding satellite image for the city\_5\_philadelphia\_28 scenario. The simulation carrier frequency is 28~GHz. A $64$-element ULA is employed at the BS with half-wavelength antenna spacing, and each channel is reconstructed from the $L_{\mathrm{path}}=5$ strongest propagation paths using \eqref{eq:channel}. To emulate temporal evolution from the static ray-tracing snapshots, consecutive points within the dataset are concatenated into trajectories where the step size between adjacent points is approximately $1.5$~m which is also used as the prediction step size. We assign a sampling interval of $\Delta t=100$~ms, corresponding to a UE speed of $v=15$~m/s. Previous studies have shown that a beam can remain useful over a beam coherence time which is much longer than the channel coherence time~\cite{Khorsandmanesh2024,Va2017}. Hence, beam updates are not needed over every instantaneous channel fluctuation. In this work, the beamforming vectors from the previous $T_h=15$ time slots are used as the history observations for each prediction.

The denoiser is trained using a linear variance schedule with $N_{\mathrm{diff}}=500$ diffusion steps, where the variance increases linearly from $\beta_1=10^{-4}$ to $\beta_{N_{\mathrm{diff}}}=2\times10^{-2}$. During inference, $K=10$ uniformly spaced diffusion levels are selected for the DDIM sampling process with $\eta=0.3$. The history encoder contains six transformer layers with an embedding dimension of $256$ and eight attention heads. The one-dimensional U-Net contains four downsampling stages, with the angular-bin length decreasing from $64$ to $4$ and the channel width increasing from $128$ to $2048$.

The model is trained for up to $100$ epochs using AdamW with a batch size of $256$, a learning rate of $2\times10^{-4}$, and a weight decay of $5\times10^{-4}$. The detailed simulation parameters are summarized in Table~\ref{tab:setup}.

\begin{table}[t]
    \centering
    \caption{Simulation parameters}
    \label{tab:setup}
    \begin{tabular}{lc}
        \toprule
        Parameter & Value \\
        \midrule
        BS antennas $N_T$ & 64 \\
        Carrier frequency $f_c$ & 28 GHz \\
        Antenna spacing $d/\lambda$ & 0.5 \\
        Retained channel paths $L_{\mathrm{path}}$ & 5 \\
        History length $T_{\rm h}$ & 15 \\
        Training diffusion steps $N_{\mathrm{diff}}$ & 500 \\
        DDIM inference steps $K$ & 10 \\
        DDIM stochasticity parameter $\eta$ & 0.3 \\
        Beta range & $10^{-4}$ to $2\times10^{-2}$ \\
        Dropout & 0.1 \\
        Batch size & 256 \\
        AdamW learning rate & $2\times10^{-4}$ \\
        AdamW weight decay & $5\times10^{-4}$ \\
        Epochs & 100 \\
        \bottomrule
    \end{tabular}
\end{table}

\subsection{Road-Aware Site-Specific Trajectory Construction}
In a site-specific environment, two different types of information can benefit beam prediction. First, the site geometry imposes spatial constraints on the feasible beam directions. Second, mobility patterns provide temporal information which is particularly important for beam prediction. For example, vehicles may follow road lanes and pedestrians may move along accessible paths, leading to structured transitions such as straight movement or turning at intersections. These mobility patterns determine how the future beam state evolves from the observed beam history, and are conceptually distinct from the instantaneous geometric constraint.
In practice, privacy concerns and the limited availability of real-world UE trajectories make it difficult to obtain representative mobility traces. We therefore construct a road-aware mobility model to generate plausible urban trajectories while respecting the local road layout. The main steps are summarized as follows.

Let $\mathcal P=\{\bm p_i\in\mathbb R^2\}_{i=1}^{N_{\rm pos}}$ denote the valid UE coordinates. We first rasterize $\mathcal P$ to form a spatial occupancy grid. Specifically, dilation and closing are performed, followed by Zhang--Suen thinning~\cite{ZhangSuen1984}, to obtain the road skeleton. The skeleton connectivity is then used to identify branch nodes and estimate the branch directions around each UE point. A point associated with at least three feasible branch directions is treated as an intersection point.
\begin{figure}[t]
	\centering
	\includegraphics[width=0.70\columnwidth]{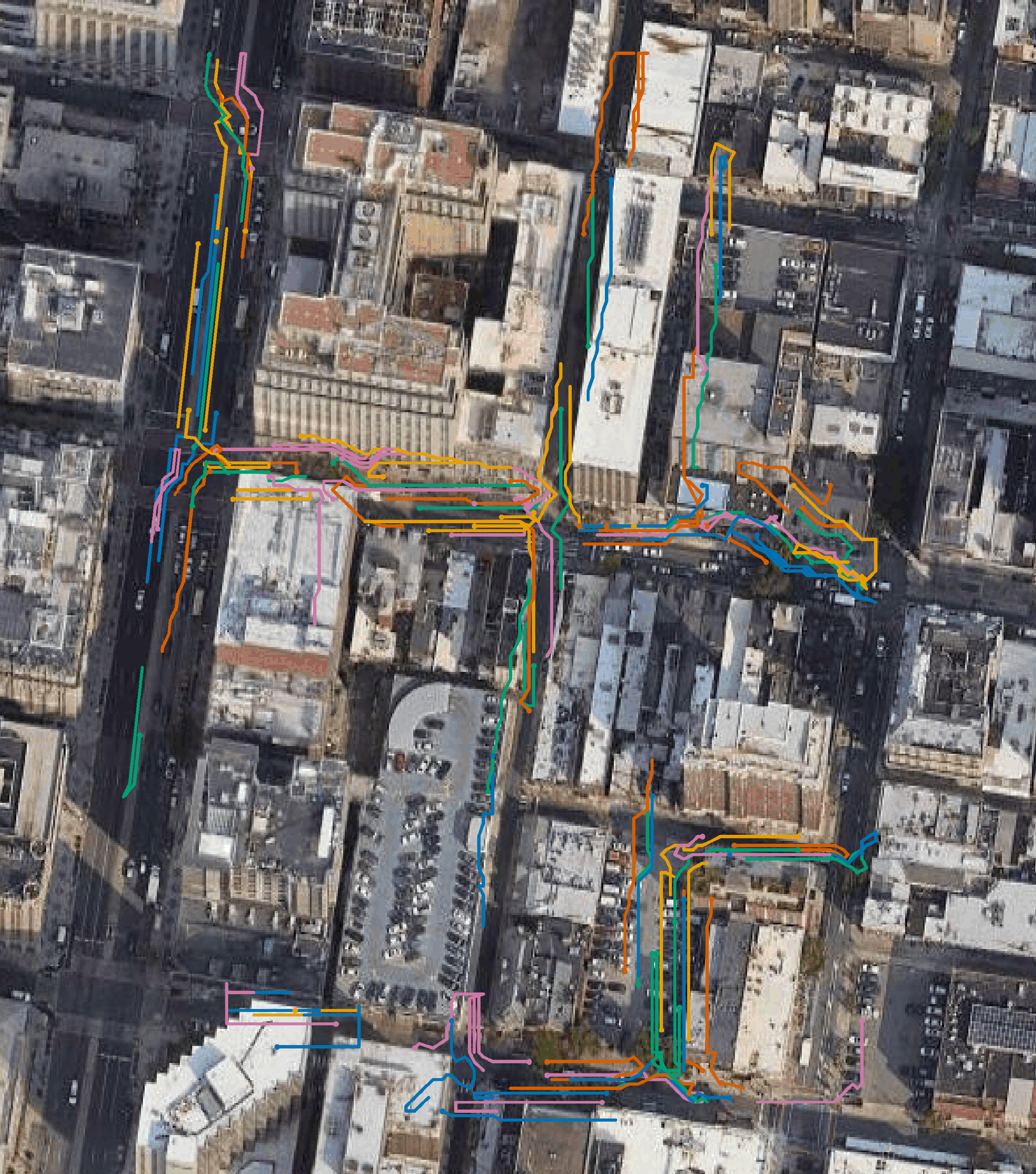}
	\caption{Illustration of road-aware trajectory construction.}
	\label{fig:track_preview}
\end{figure}

To initialize a trajectory, the starting point is sampled according to the local point density. Specifically, let $\mathcal N_i$ denote the local nearest-neighbor set of point $i$ and assign point $i$ the local-density weight
\begin{equation}
\nu_i=\left(\frac{1}{|\mathcal N_i|}\sum_{r\in\mathcal N_i}
\|\bm p_r-\bm p_i\|_2\right)^{-1},
\label{eq:road_density}
\end{equation}
where $r$ indexes the neighbors of point $i$. A smaller average neighbor distance indicates a denser region and results in a larger value of $\nu_i$. The trajectory starting point is then sampled according to the normalized weights:
\begin{equation}
\Pr(i_0=i)=\frac{\nu_i}{\sum_{n=1}^{N_{\rm pos}}\nu_n},
\label{eq:road_start}
\end{equation}
where $i_0$ denotes the starting-point index, $N_{\rm pos}$ is the total number of valid UE positions, and $n$ is a dummy index over all valid positions. This normalization gives a valid starting-point distribution. After the starting point is selected, the initial heading is set to the dominant local branch direction at that point. At step $t$, the unvisited candidate set is
\begin{equation}
\mathcal C_t^{(0)}=\left\{r\in\mathcal N_{i_t}\setminus\mathcal V_t:0<\|\bm p_r-\bm p_{i_t}\|_2\leq d_{\max}\right\},
\label{eq:road_candidates}
\end{equation}
where $\mathcal V_t$ contains the previously visited points and $d_{\max}$ is the maximum admissible step size. At a non-intersection point, the desired heading $\vartheta_t$ is the local branch direction closest to the current heading $\psi_t$. At an intersection, when turning is allowed, an action
$a_t\in\{\mathrm F,\mathrm B,\mathrm L,\mathrm R\}$ is sampled according to
the categorical prior
$\bm\pi_{\rm turn}=[p_{\rm F},p_{\rm B},p_{\rm L},p_{\rm R}]$.
The four actions represent continuing straight, making a U-turn, turning left,
and turning right, corresponding to heading offsets
$(0,\pi,\pi/2,-\pi/2)$ relative to the current heading $\psi_t$, respectively. Candidate points sufficiently aligned with $\vartheta_t$ are retained, and the set is further restricted to points near the target step size $d_0$. We denote the resulting candidate set by $\mathcal C_t$. The next point is then sampled according to
\begin{subequations}
\label{eq:road_transition}
\begin{align}
\Pr(i_{t+1}=r\mid i_t)
&=\frac{\exp(\chi_{t,r}/\kappa)}{\sum_{n\in\mathcal C_t}\exp(\chi_{t,n}/\kappa)},
\label{eq:road_transition_prob}\\
\chi_{t,r}
&=-\frac{|d_{t,r}-d_0|}{\delta_d}
-\omega_\theta\frac{|\operatorname{wrap}(\psi_{t,r}-\vartheta_t)|}{\delta_\theta}.
\label{eq:road_transition_score}
\end{align}
\end{subequations}
where $d_{t,r}=\|\bm p_r-\bm p_{i_t}\|_2$, $\psi_{t,r}$ is the candidate heading, $\delta_d$ and $\delta_\theta$ control the step-size and heading tolerances, respectively, $\omega_\theta$ weights directional consistency, and $\kappa$ is the sampling temperature. Notably, after a sufficiently large heading change, a turn cooldown is imposed to avoid successive unrealistic turns. Sampling continues until the prescribed trajectory length is reached. The coordinates are then mapped to their corresponding channel and beam states, and consecutive labels form the beam-history and target pairs $(\mathcal H_t,\bm w_{t+1}^\star)$. A subset of the trajectories constructed by this road-aware mobility model is illustrated in Fig.~\ref{fig:track_preview}.

\begin{figure}[t]
	\centering
	\includegraphics[width=0.9\columnwidth]{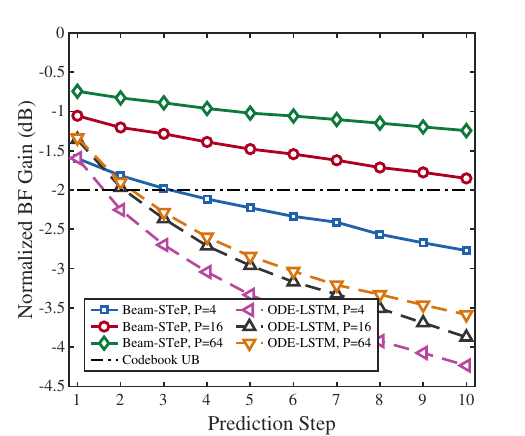}
	\caption{Normalized beamforming gain performance achieved by different schemes in the City\_5\_philadelphia\_28 scenario.}
	\label{fig:philadelphia_rollout}
\end{figure}

\begin{figure}[t]
	\centering
	\includegraphics[width=0.9\columnwidth]{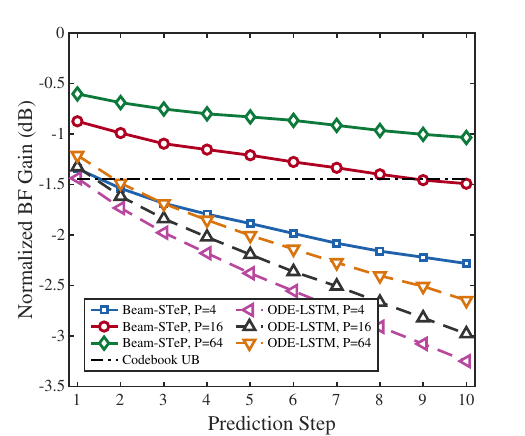}
	\caption{Normalized beamforming gain performance achieved by different schemes in the City\_17\_seattle\_28 scenario.}
	\label{fig:seattle_rollout}
\end{figure}

\begin{figure}[t]
	\centering
	\includegraphics[width=0.9\columnwidth]{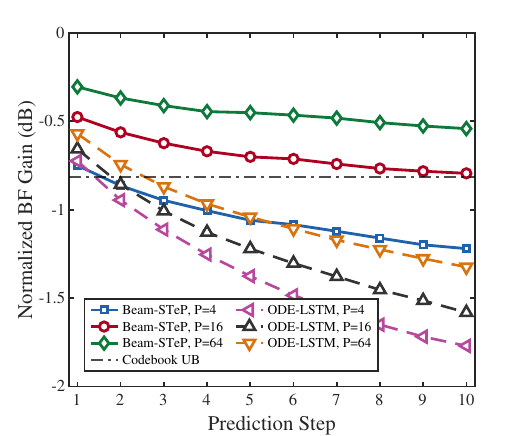}
	\caption{Normalized beamforming gain performance achieved by different schemes in the City\_2\_chicago\_28 scenario.}
	\label{fig:chicago_rollout}
\end{figure}

After the construction, the trajectories, rather than individual points, are partitioned into training, validation, and test sets with a ratio of 70\%, 15\% and 15\%. Each scenario contains 18000 trajectories. Here, after trajectory-level splitting, we verify that the training and test sets contain no identical history-target windows, thereby preventing exact-window data leakage.

\begin{figure*}[t]
	\centering
	\includegraphics[width=0.9\textwidth]{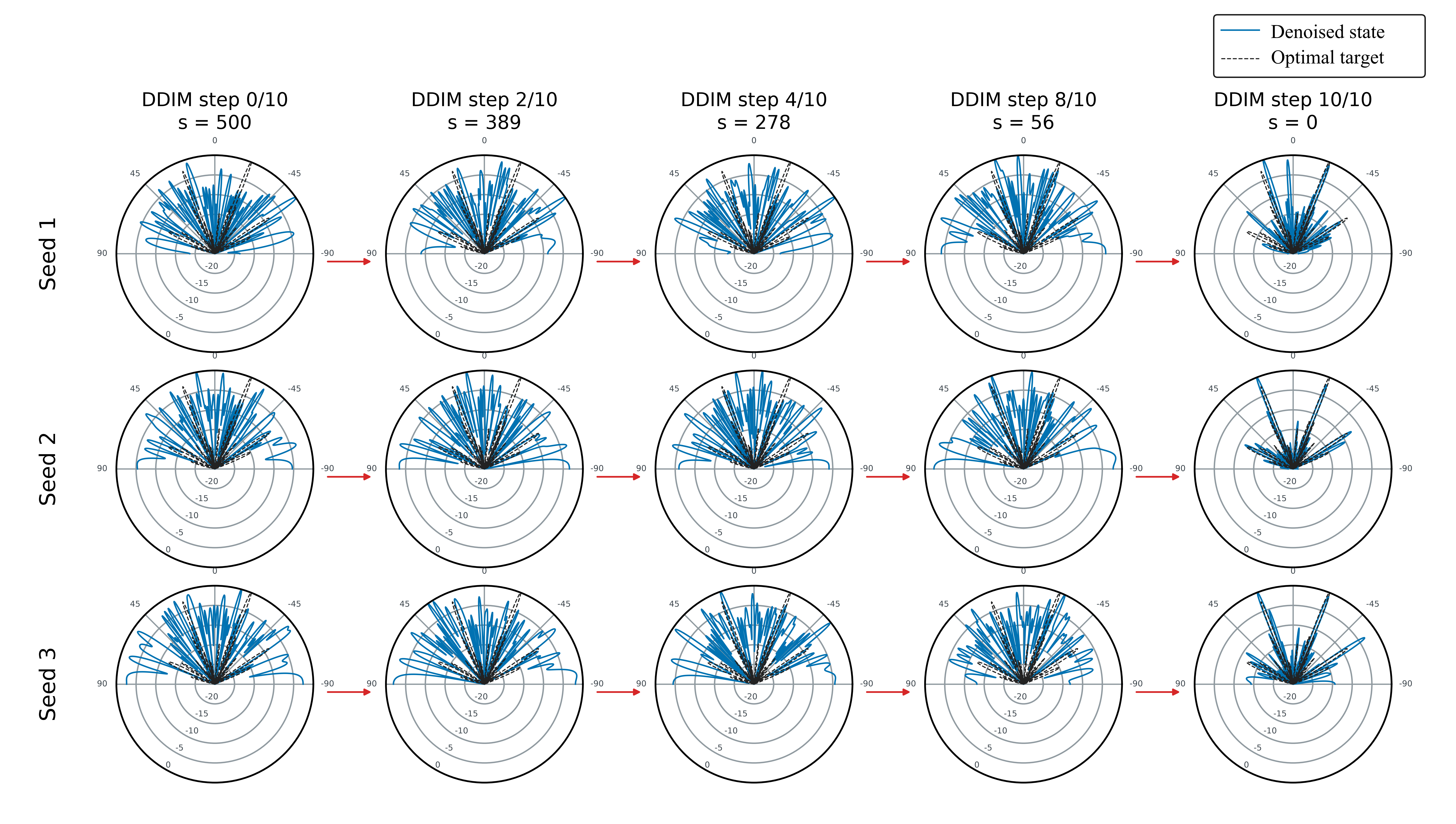}
	\caption{Selected denoising process of an intersection area sample in the City\_5\_philadelphia\_28 scenario.}
	\label{fig:denoising_process}
\end{figure*}

\begin{remark}This road-aware mobility model is not intended to reproduce the exact statistical behavior of real users. Instead, it provides a controllable approximation for studying how site-specific mobility-pattern information contributes to beam prediction problem. Therefore, other mobility models may also be adopted to evaluate the proposed method.
\end{remark}

\begin{table}[t]
	\centering
	\caption{Comparison of adjacent-slot last-beam reuse and one-step Beam-STeP prediction under different probing budgets}
	\label{tab:last_beam_variation}
	\footnotesize
	\setlength{\tabcolsep}{4pt}
	\begin{tabular}{@{}lcccc@{}}
		\toprule
		Scenario & \shortstack{Last Beam \\ (dB)} & \shortstack{Beam-STeP \\ $P=4$ (dB)} & \shortstack{Beam-STeP \\ $P=16$ (dB)} & \shortstack{Beam-STeP \\ $P=64$ (dB)} \\
		\midrule
		Philadelphia & $-2.405$ & $-1.598$ & $-1.055$ & $-0.743$ \\
		Seattle & $-1.977$ & $-1.344$ & $-0.875$ & $-0.604$ \\
		Chicago & $-1.228$ & $-0.750$ & $-0.476$ & $-0.305$ \\
		\bottomrule
	\end{tabular}
\end{table}

\subsection{Baselines}
The following baselines are considered for performance comparison.
\subsubsection{Last-Beam Reuse}
As a simple non-predictive baseline, the last-beam strategy directly reuses the optimal beamformer from the preceding slot for the current slot. Its performance quantifies whether the beam variation between two adjacent prediction slots is sufficiently large to require predictive beam updating.
\subsubsection{Continuous Multi-Candidate ODE-LSTM}
For discriminative beam prediction methods, LSTM-based architectures are widely adopted. Inspired by \cite{Ma2023a}, we use an ODE-LSTM model as a representative discriminative baseline. At each slot, a multilayer perceptron projects the historical continuous angular-domain BF vector into a feature embedding, and the resulting embedding sequence is fed to a one-layer LSTM. Two fully connected ODE layers then evolve the final LSTM hidden state. To match the prediction setting of Beam-STeP, we adapt ODE-LSTM
to take historical beamforming (BF) vectors sampled at discrete
time slots as input. A candidate head predicts multiple
continuous-valued BF vectors in the angular domain, while a score
head assigns a confidence logit to each candidate. Softmax
normalization then converts these logits into probabilities for
selecting the candidates to be probed.
\subsubsection{Codebook-Based Method}
Codebook-based beam prediction is a widely used low-complexity approach for practical beam management. Previous studies have reported promising performance with codebook-based methods, e.g., \cite{Azarbahram2026}. However, many of these methods rely on auxiliary observations, such as RSRP measurements collected from previous beam probes. Incorporating such information would lead to a different observation model and probing budget from those considered in this paper, making a direct comparison less fair. Therefore, instead of implementing a particular codebook-based predictor, we report the 64-beam codebook upper bound. At each time slot, this bound selects the codeword that achieves the maximum beamforming gain for the realized channel. Notably, practical codebook-based predictors generally remain below this bound because of imperfect beam prediction and limited observations.
\begin{figure*}[t]
	\centering
	\subfloat[City\_5\_philadelphia\_28\label{fig:site_ablation_philadelphia}]{%
		\includegraphics[width=0.325\textwidth]{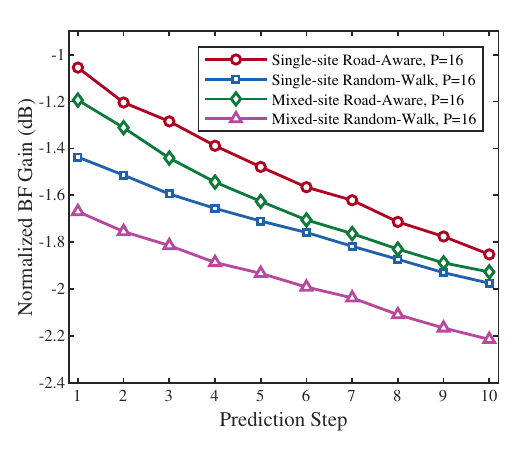}}%
	\hfill
	\subfloat[City\_17\_seattle\_28\label{fig:site_ablation_seattle}]{%
		\includegraphics[width=0.325\textwidth]{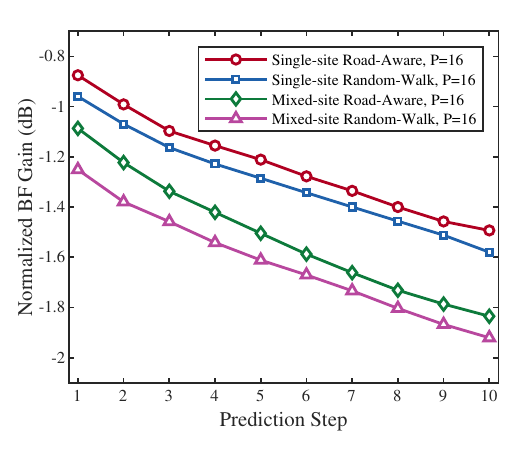}}%
	\hfill
	\subfloat[City\_2\_chicago\_28\label{fig:site_ablation_chicago}]{%
		\includegraphics[width=0.325\textwidth]{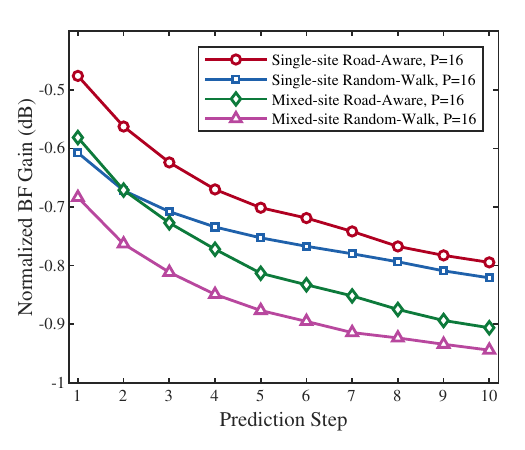}}
	\caption{Performance comparison of Beam-STeP under different site-specific training setups in the (a) City\_5\_philadelphia\_28, (b) City\_17\_seattle\_28, and (c) City\_2\_chicago\_28 scenarios.}
	\label{fig:site_ablation}
\end{figure*}
\subsubsection{Site-Agnostic Methods}
To isolate the importance of site-specific propagation and mobility knowledge, we construct three controlled variants while keeping the network architecture, training budget, and probing protocol unchanged. The \emph{single-site random-walk} variant retains channels from the target site but replaces road-aware trajectories with unconstrained random walks, thereby removing site-specific mobility patterns. The \emph{mixed-site road-aware} variant uses equally sampled training data from all three sites while retaining road-aware trajectories, thus weakening the target-site propagation prior and representing a general multi-site training. Finally, the \emph{mixed-site random-walk} variant combines mixed-site channels with random-walk mobility, removing both sources of site-specific knowledge.

\subsection{Performance of Beamforming Gain}
The reported performance averages the linear normalized gain over $N$ test samples,
\begin{equation}
	G_{\rm norm}^{(P)}=10\log_{10}\left(
	\frac{1}{N}\sum_{i=1}^{N}
	\frac{\max_{m\leq P}g_i(\widehat{\bm w}_i^{(m)})}
	{g_i(\bm w_i^\star)}\right).
	\label{eq:metric}
\end{equation}
The gain is evaluated after inverse DFT and the constant-modulus projection. The probing budgets reported below are $P\in\{4,16,64\}$.

Table~\ref{tab:last_beam_variation} reports the performance of last-beam reuse at the first prediction slot, together with that of Beam-STeP under different probing budgets. Directly reusing the optimal beam from the preceding slot results in normalized gains of $-2.405$~dB, $-1.977$~dB, and $-1.228$~dB in city\_5\_philadelphia\_28, city\_17\_seattle\_28, and city\_2\_chicago\_28 scenarios, respectively. With only $P=4$ candidates, Beam-STeP improves these results by $0.81$~dB, $0.63$~dB, and $0.48$~dB, demonstrating that the beam variation between adjacent slots is non-negligible and highlighting the necessity of beam prediction even over a single-slot horizon. Further gains are obtained as the probing budget increases, indicating that multiple generated candidates provide better coverage of the possible next-beam states.

Figs.~\ref{fig:philadelphia_rollout} -- \ref{fig:chicago_rollout} present the multi-step beam prediction performance of the proposed method in the three scenarios. Beam-STeP is compared with the discriminative ODE-LSTM baseline and the upper bound of codebook-based methods. Under the same beam probing overhead, Beam-STeP consistently achieves higher beamforming gains than the ODE-LSTM baseline. This advantage is reflected both in the higher gain obtained at each prediction step and in the slower degradation as the prediction horizon increases.

This improvement can be attributed to the multimodal nature of future beam evolution. Specifically, user mobility and wireless propagation jointly induce multiple plausible future beam patterns. Our generative model can learn this conditional distribution and generate diverse candidates under the same historical observation, allowing the probing process to cover a wider range of possible future beam directions. This candidate diversity reduces the cumulative prediction error and leads to more stable multi-step performance. As shown in the figures, with a probing budget of $P=64$, Beam-STeP outperforms the ODE-LSTM baseline at the tenth prediction step by approximately $2.34$~dB, $1.61$~dB, and $0.78$~dB in the city\_5\_philadelphia\_28, city\_17\_seattle\_28, and city\_2\_chicago\_28 scenarios, respectively. The evolution of the gain across prediction steps further illustrates the difference in recursive prediction performance. Under the same probing budget of $P=64$, the normalized beamforming gain of Beam-STeP decreases by approximately $0.50$~dB, $0.43$~dB, and $0.24$~dB from the first to the tenth prediction step in the three scenarios. In comparison, ODE-LSTM exhibits larger decreases of $2.25$~dB, $1.44$~dB, and $0.75$~dB. Since the beam selected at each step is appended to the history for subsequent predictions, an inaccurate selection can affect both current transmission and future prediction conditions. The smaller gain degradation observed for Beam-STeP is consistent with improved robustness to error propagation during recursive prediction. Moreover, since the generated candidates provide stronger diversity, increasing the probing budget leads to a larger performance improvement for Beam-STeP than for ODE-LSTM. This trend suggests that the diffusion-generated candidates provide less redundant coverage of the conditional future-beam distribution.

For the codebook-based baseline, we report an upper bound that selects the best beam from all 64 fixed codewords for each realized channel. Therefore, the comparison at $P=64$ ensures the same beam-probing overhead for both methods. The results show that Beam-STeP outperforms the codebook upper bound in all three scenarios. Notably, with $P=16$, Beam-STeP uses only one quarter of the probing budget required for exhaustive sweeping of the 64-beam DFT codebook. It exceeds the corresponding codebook upper bound throughout the ten-step prediction horizon in Philadelphia and Chicago, and over
the first eight prediction steps in Seattle, with only a small gap below the reference at the final two steps. This advantage mainly results from the inherent grid mismatch and beam quantization of codebook-based beam selection.

\begin{table*}[t]
	\centering
	\caption{Candidate-generation latency and normalized beamforming gain on an NVIDIA A100-SXM4-80GB GPU ($P=16$, $h=10$).}
	\label{tab:inference_latency_comparison}
	\begingroup
	\footnotesize
	\setlength{\tabcolsep}{7pt}
	\renewcommand{\arraystretch}{1.15}
	\begin{tabular}{lcrrrrrr}
		\toprule
		\multirow{2}{*}{Sampler} & \multirow{2}{*}{Steps}
		& \multicolumn{2}{c}{Philadelphia}
		& \multicolumn{2}{c}{Seattle}
		& \multicolumn{2}{c}{Chicago} \\
		\cmidrule(lr){3-4} \cmidrule(lr){5-6} \cmidrule(lr){7-8}
		& & Latency (ms) & Gain (dB) & Latency (ms) & Gain (dB)
		& Latency (ms) & Gain (dB) \\
		\midrule
		DDIM & 5   & 20.85   & $-2.260$ & 21.22   & $-1.917$ & 21.40   & $-0.964$ \\
		\textbf{DDIM} & \textbf{10}
		& \textbf{40.45} & $\bm{-1.852}$
		& \textbf{40.27} & $\bm{-1.493}$
		& \textbf{42.71} & $\bm{-0.794}$ \\
		DDIM & 50  & 184.41  & $-1.718$ & 191.17  & $-1.418$ & 193.08  & $-0.740$ \\
		\midrule
		DDPM & 500 & 2882.63 & $-1.819$ & 2505.23 & $-1.417$ & 2625.38 & $-0.740$ \\
		\bottomrule
	\end{tabular}
	\par\smallskip
	\parbox{\textwidth}{\footnotesize
	}
	\endgroup
\end{table*}

To further illustrate the multimodal nature of Beam-STeP, Fig.~\ref{fig:denoising_process} visualizes the denoising processes of multiple predicted candidates for an intersection sample from the city\_5\_philadelphia\_28 scenario. For a UE located near an intersection, multiple future heading directions may be possible, including continuing straight or turning onto another branch. These mobility alternatives can lead to distinct future beam directions. Conditioned on the same historical beamforming vector sequence, three independent Gaussian noise samples are initialized and progressively denoised by the network. The samples converge to different final beam states with distinct main lobe directions, representing several plausible future beam transition possibilities at the intersection area. In particular, the beam generated with Seed 2 closely matches the target beam. The resulting patterns also preserve rich sidelobe structures and even multiple main lobes rather than being characterized only by a single main lobe direction. This highlights another advantage over codebook-based methods: by generating continuous beamforming vectors instead of selecting from a fixed set of quantized beams, the proposed Beam-STeP can reproduce the complete beam pattern and more closely approximate the continuous optimal beam. 

Notably, although only one candidate closely matches the realized target in this example, the remaining candidates represent alternative hypotheses under the same history observations. Online probing resolves this uncertainty by selecting the candidate with the highest measured RSRP.

\subsection{Site-Specific Analysis}
Fig.~\ref{fig:site_ablation} compares our Beam-STeP method with three other site-agnostic conditions across the three scenarios. The single-site road-aware model is the full site-specific reference. The single-site random-walk model retains the site-specific propagation data but removes the road-constrained mobility pattern. The mixed-site road-aware model combines data from the three scenarios and therefore weakens the single-site propagation prior while retaining road-aware mobility. Finally, the mixed-site random-walk model removes both the single-site propagation prior and the road-aware mobility pattern. All four models use the same $P=16$ probing budget.

The numerical results show that the fully site-specific approach, trained with single-site propagation data and road-aware trajectories, consistently outperforms both the single-site random-walk model and the mixed-site road-aware model in all three scenarios. The improvement over the single-site random-walk model indicates that road-aware mobility patterns provide useful temporal information for predicting future beam states. Meanwhile, the improvement over the mixed-site road-aware model demonstrates the value of site-specific propagation data in constraining the conditional beam distribution. These gains remain consistent as the prediction horizon increases, showing that site-specific information is beneficial not only for short-term prediction but also for recursive multi-step beam tracking. Moreover, the mixed-site random-walk model generally achieves the lowest performance, confirming that propagation and mobility information provide complementary benefits. Since all models use the same architecture, training settings, and probing protocol, the performance differences can be attributed to the site-specific information used during data construction. This consistent trend across the three scenarios also demonstrates the robustness of the proposed site-specific design under different propagation geometries.
\begin{figure}[t]
	\centering
	\includegraphics[width=0.75\columnwidth]{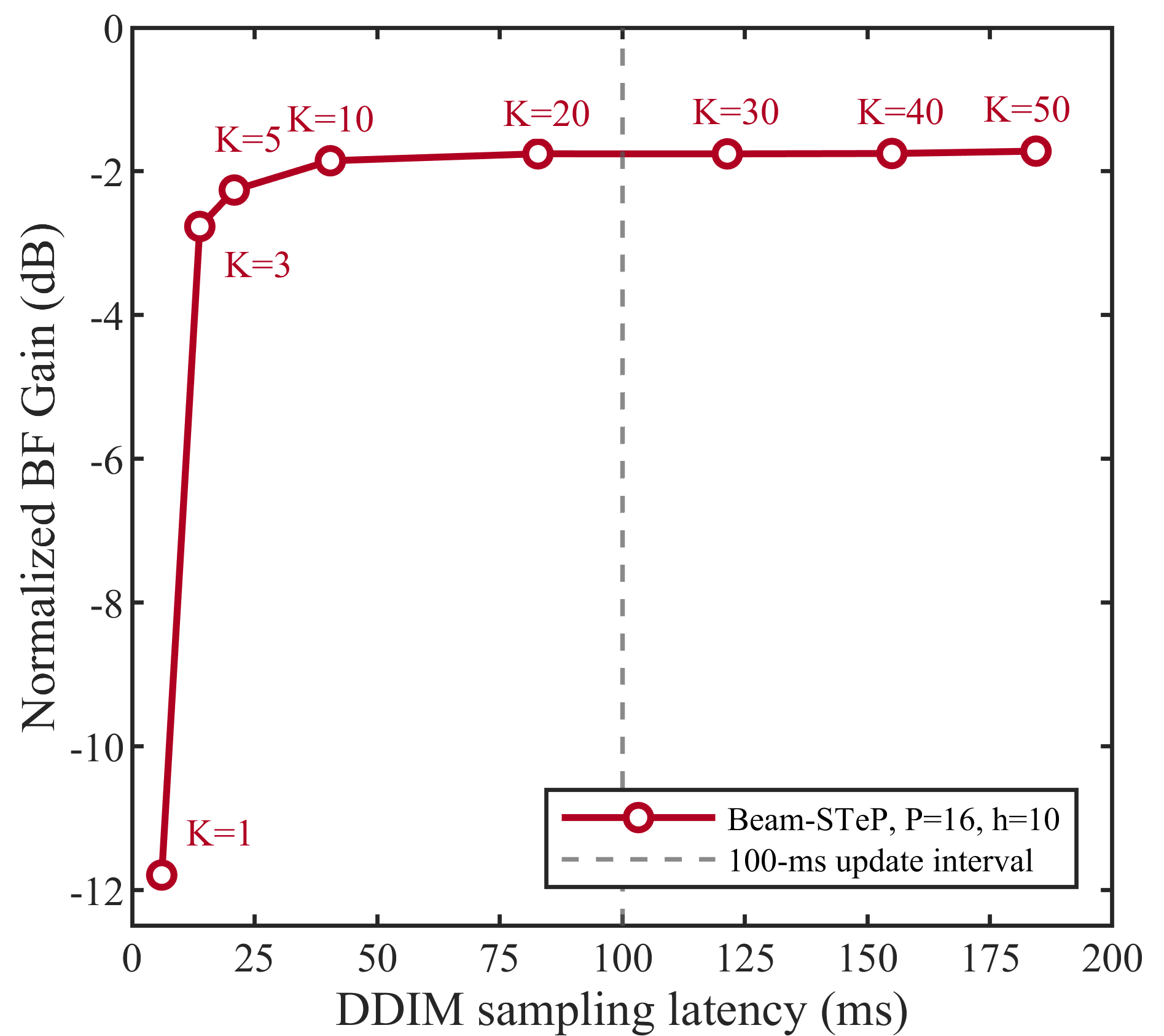}
	\caption{Gain-latency trade-off in the city\_5\_philadelphia\_28 scenario at $P=16$ and $h=10$, with $K$ annotated.}
	\label{fig:latency}
\end{figure}

\subsection{Online Inference Latency Analysis}

Fig.~\ref{fig:latency} and
Table~\ref{tab:inference_latency_comparison} characterize the
performance--latency trade-off of the proposed sampling procedure.
The reported latency is measured on an NVIDIA A100-SXM4-80GB GPU for
generating a batch of $P=16$ beam candidates at one prediction step.
Since the same network and sampling procedure are invoked at each
recursive prediction step, the candidate-generation latency is
approximately independent of the prediction horizon. Moreover, the
$P$ candidates are processed in parallel as a batch on the GPU.
Accordingly, the normalized beamforming gain at the tenth prediction
step ($h=10$) is reported as a representative long-horizon result.

Fig.~\ref{fig:latency} provides a detailed comparison for the
Philadelphia scenario. With only one DDIM sampling step, candidate
generation requires 6.05~ms, but the normalized beamforming gain drops
to $-11.79$~dB because of insufficient denoising. Increasing the number
of sampling steps to $K=10$ substantially improves the gain to
$-1.85$~dB at a latency of 40.45~ms. Beyond this point, additional
sampling steps yield diminishing performance improvements while the
latency continues to increase approximately linearly. In particular,
$K=20$ achieves $-1.76$~dB at 82.81~ms, whereas $K=50$ requires
184.41~ms for only a further 0.04~dB improvement. Therefore, $K=10$
reduces the candidate-generation latency by 78.1\% relative to $K=50$,
with a gain reduction of only 0.13~dB.

Table~\ref{tab:inference_latency_comparison} further shows that this
trade-off is consistent across the three considered scenarios. At
$K=10$, candidate generation requires around 40~ms across the three scenarios, and the
resulting beamforming gain remains within 0.13~dB of that achieved by
the 500-step DDPM sampler. Compared with DDPM, the 10-step DDIM sampler
reduces the average latency by approximately 98.5\%, corresponding to an
approximately 65 times speedup. Increasing $K$ from 10 to 50 raises the
latency by approximately 4.6 times, while improving the gain
by only around 0.1~dB. At $K=50$, DDIM achieves essentially the same
gain as DDPM but remains over 10 times faster. The small
site-to-site variation in latency also indicates that the online
computational cost is governed primarily by the sampler configuration
rather than the propagation environment. These results identify
$K=10$ as a favorable operating point. The candidate-generation
latency of $K=10$ is below the assumed update interval of $\Delta t=100$~ms
across all three scenarios, while retaining performance close to that
of substantially longer sampling procedures. Nevertheless, the reported values account only for candidate generation. The overall end-to-end latency also includes candidate
postprocessing, beam probing, RSRP feedback, and beam selection.

\section{Conclusion}
In this paper, we proposed Beam-STeP, a generative site-specific model for beam prediction. Beam-STeP jointly exploits spatial intelligence from site-dependent propagation geometry
and mobility patterns, together with temporal intelligence
from historical beamforming sequences. A history-conditioned diffusion model with a DDIM sampler is deployed to generate
multiple feasible future beam candidates, whose diversity enables effective beam
probing and selection. Simulation results across three ray-tracing scenarios
demonstrated that Beam-STeP generally outperforms the considered baselines and maintains robust multi-step
prediction performance, with larger probing budgets effectively mitigating performance degradation over recursive prediction horizons. Moreover, the latency evaluation demonstrated a flexible trade-off between beamforming gain and inference latency, with the 10-step DDIM sampler maintaining near-DDPM performance while keeping the candidate-generation latency within a practical range.

\balance

\end{document}